**Magnetic and magnetotransport behavior of RFe$_5$Al$_7$ (R= Gd and Dy): Observation of 're-entrant inverse-magnetocaloric phenomenon' and asymmetric magnetoresistance behavior**

Venkatesh Chandragiri, Kartik K Iyer and E.V. Sampathkumaran*
*Tata Institute of Fundamental Research, Homi Bhabha Road, Colaba, Mumbai 400005, India*

**Abstract**

We have compared and contrasted magnetic, magnetocaloric and magnetoresistive properties of Gd and Dy members of the rare-earth (R) series RFe$_5$Al$_7$, crystallizing in ThMn$_{12}$ structure, known to order antiferromagnetically. Among other observations, we would like to emphasize on the following novel findings: (i) There are multiple sign-crossovers in the temperature ($T$) dependence of isothermal entropy change ($\Delta S$) in the case of Dy compound; in addition to nil $\Delta S$ at the magnetic compensation point known for two-magnetic-sublattice systems, there is an additional sign-crossover at low temperatures, as though there is a 're-entrant inverse magnetocaloric' phenomenon. Corresponding sign reversals could also be observed in the magnetoresistance data. (ii) The plots of magnetoresistance versus magnetic field are found to be highly asymmetric with the reversal of the direction of magnetic-field ($H$) well below $T_N$ for both compounds, similar to that known for an antiferromagnetic tunnel junctions. We attribute these to subtle changes in spin orientations of R and Fe moments induced by $T$ and $H$.

# I. INTRODUCTION

The magnetocaloric effect (MCE) of rare-earth based ferrimagnets containing two magnetic sublattices, namely rare-earth (R) based garnets, attracted special interest a few decades ago, as there is a sign cross-over at the magnetic compensation point [1]. This sign crossover is due to the fact that each of the antiferromagnetically coupled sublattices contributes to MCE. This issue for three magnetic sub-lattice systems was also theoretically addressed successfully considering a model Hamiltonian [2] that includes the exchange interaction terms between the inter sublattices, and intra sublattices as well as a Zeeman interaction term. We consider it important to further expand our knowledge on MCE phenomenon from this angle by focusing on other multi-sublattice magnetic systems.

In this respect, the rare-earth intermetallics of the type, $RFe_5Al_7$, crystallizing in $ThMn_{12}$-type tetragonal (*I4/mmm*) structure [3, 4] provide an ideal opportunity to probe the same. At the time of early investigations of these compounds [3], it was proposed that this family of materials displays rich magnetic phenomena, like huge thermal and magnetic hysteresis, first observation of large negative magnetization and magnetic compensation points. In fact, the interest in this family seems to have revived only very recently [5, 9] after a few decades of initial work in early 1980s. In this family, R, Fe and Al atoms occupy the 2a, 8f sites and 8i sites respectively; in addition, 8j site is shared by Fe and Al. It is known that the heavy rare-earth members order ferrimagnetically at rather high temperatures ($T_c$ = 200 to 270 K) [3, 10, 11] with 2a-2a and 8j-8j intra-sublattice exchange interactions being ferromagnetic and inter-sublattice interaction being antiferromagnetic. It appears that Fe at 8i site does not order magnetically, while the exact nature of magnetic ordering of Fe at 8f site is not clear [3, 4, 10]. With the primary motivation of understanding of MCE behavior of such a complex ferrimagnet, we have undertaken a magnetic study to compare and contrast the behavior of two members, namely R= Gd and Dy, of this family. These two members order ferrimagnetically at 250 and 223 K respectively; while magnetic compensation point ($T_0$) occurs around 92 K for the Dy compound, it is believed that full compensation is not attained down to 4 K for the Gd compound [5, 7]. Strong magnetic anisotropy has been observed with [100] and [110] axes being the easy axis for Dy and Gd members respectively, based on single crystal magnetic studies [5, 7].

# II. EXPERIMENTAL DETAILS

The compounds in the polycrystalline form were prepared by arc-melting stoichiometric amounts of constituent elements (Gd 99.99%, Dy 99.99%, Fe 99.99% and Al 99.999%) in an atmosphere of argon. The weight loss after final melting was lower than 0.8%. The samples thus prepared were characterized by x-ray diffraction (Cu $K_\alpha$) and found to be single phase. Rietveld refinements were carried out on the observed patterns and the lattice constants and other parameters (see figure 1) compare quite well with those reported in the literature [4 - 6]. The temperature (*T*) dependence of magnetization (*M*) as well as isothermal *M* measurements as a function of magnetic field (*H*) were carried out using commercial magnetometers (Quantum Design, USA). Heat-capacity (*C*) as a function of *T* (also in the presence of some fixed magnetic fields) were carried out by a commercial Physical properties Measurements System (PPMS) (Quantum Design, USA). The same PPMS was used to study electrical resistivity (ρ) and magnetoresistance (MR) behavior as a function of *T* and *H* by four-probe method using silver paint for making electrical contacts of the leads with the sample; *H* was applied perpendicular to the direction of the excitation current. Unless otherwise stated, all measurements were carried out by cooling the samples to desired temperatures from much above their respective Curie temperatures.

# III. RESULTS AND DISCUSSION
## A. Magnetic susceptibility

Though magnetization behavior is known in the literature, for the sake of completeness, we first discuss the temperature dependence of magnetization, *M(T)*, measured under the applied magnetic field of 100 Oe and 5 kOe (see figure 2) for both the samples. For Gd sample, for $H$= 100 Oe, the zero-field-cooled (ZFC) curve exhibits a sharp upturn around 255 K with decreasing $T$, followed by a broad peak around 200 K; similar behavior is seen in the curve taken with 5 kOe, however revealing a broadening of the feature near $T_C$ (see Fig. 2a and Fig. 2b). The field-cooled (FC) curves exhibit similar behavior, but the one measured with 100 Oe separates from the respective ZFC-curve below about 170 K. It should be noted that there is no evidence for zero magnetization in these curves, with the values of $M$ remaining in the positive zone. Now, turning to the Dy sample, there is a sharp change at $T_c$ (see Fig. 2c); ZFC-FC curves obtained with 100 Oe separate below about 210 K, as in the case of Gd, but this separation appears in the 5kOe-curves (unlike in Gd sample) at the low temperature region below about 45 K (see Fig. 2d). Major differences with respect to the behavior in the Gd compound are: (i) There is a negative magnetization in the ZFC-curve below 92 K, if measured with low external fields (e.g., 100 Oe); (ii) there is a magnetic compensation point around 92 K in the curve, if measured with 5 kOe. However, clearly, a complete compensation of antiferromagnetically coupled Gd moments and Fe moments does not occur in the Gd alloy, because $M_{Fe}$ is greater than $M_{Gd}$ [5]. Additional feature worth stressing is that there is a peak in the 5kOe-ZFC curve around 45 K as shown in Fig. 2d, signaling a change in the relative alignment of Dy and Fe moments. All these findings are in broad agreement with the data reported in the literature.

### B. Isothermal entropy change behavior

We now discuss the isothermal $M$ behavior with the primary aim of understanding MCE. For this purpose, we have obtained the virgin $M(H)$ curves up to 50 kOe at close intervals of temperature (2 to 3 K) by cooling the specimens in zero-field to desired temperatures. Representative $M(H)$ curves are shown in figures 3a and 3b. For both samples, there is a step at low fields at low temperatures (near 0.5 and 25 kOe for Gd and Dy compounds respectively at 2 K, see also Section III.D), which gradually vanishes with increasing temperature. It is possible that these are due to magnetic-field-induced spin-reorientations. This feature however was apparently missed out in the past literature. The isothermal entropy change, $\Delta S$ [= $S(H_2)$-$S(H_1)$], for a change of the magnetic field from $H_1$ to $H_2$ which is a measure of *MCE*, was determined from

$$\Delta S = \int_{H_1}^{H_2} \left(\frac{\partial M}{\partial T}\right)_H dH$$

on the basis of Maxwell relationship

$$\left(\frac{\partial S}{\partial H}\right)_T = \left(\frac{\partial M}{\partial T}\right)_H$$

The values of -$\Delta S$ as a function of temperature thus derived from these $M(H)$ curves are plotted in figure 4a and 4b for Gd and Dy compounds respectively for changes of selected magnetic-fields (with $H_1 = 0$). The values are rather small in the entire temperature range of investigation. In the case of Gd compound, there is a peak in -$\Delta S$ at $T_C$, the sign of which is positive, consistent with net ferromagnetism [12]. With decreasing temperature, the sign of –$\Delta S$ gets changed near a characteristic temperature ($T^*$, Ref. 2) of 210 K, establishing that the entropy increases, rather than showing a decrease, with increasing $H$, exhibiting a broad peak around 90 K. This entry to 'inverse MCE' region persists down to very low temperatures. Below about 12 K, there is a tendency for the –$\Delta S$ curve to re-enter positive zone, which is illustrated by plotting the data as a function of $H$ in the inset-I of figure 4a. It is possible that this sign-crossover at low temperatures is a consequence of the tendency for the compensation of the magnetic moments of Gd and Fe. We noted that the observation of this feature subtly depends on various factors, e.g., the history of measurements, which influence relative orientations of the moments of Gd and Fe at very low temperatures. Presumably due to this reason, this was not detected in the past literature. In the case of Dy sample, as in Gd sample, with a lowering of $T$, a peak in the positive zone at $T_C$, a sign-crossover around about 190 K ($T^*$) followed by another broad peak around 120 K, and another sign crossover near $T_0$ could be observed. In addition, there is another peak near 70 K below $T_0$. It is interesting that this finding coupled with signatures at $T_C$, $T^*$ and $T_0$ theoretically established for three sublattice models [2] for garnets are reproduceable for intermetallic situation with more than one magnetic sublattice and with intrinsic crystallographic disorder effects. A fascinating observation, we would like to stress is that this compound is characterized by an additional change of sign near 40 K as shown in inset-I of figure 4b (followed by another broad peak), as though there is a new characteristic temperature due to reentry into another 'inverse MCE regime'. This temperature may be correlated to the ZFC-FC curves bifurcation temperature in the $M(T)$ curve measured with 5 kOe (see figure 2d). We speculate that this is due to subtle changes in magnetism, for example, in the orientations of the magnetic moments of Dy and Fe, induced by the application of magnetic-field in such a way that entropy increases. This tendency is more prominent with a further lowering of $T$ at very high fields. To demonstrate this, we plot the $-\Delta S(H)$ curves below 40 K for selected temperatures in the inset-II of figure 4b, in which prominent upturns due to $H$-induced magnetic transitions and sign cross-over are apparent; at these temperatures, the high-field behavior is of 'inverse MCE'-type and the field at which the transformation from 'usual' MCE to 'inverse MCE' takes place increases with decreasing temperature (see inset of figure 5b). Clearly, this compound presents a rare situation in which multiple sign crossovers (and hence 'reentrant MCE phenomenon'), presumably due to subtle changes in relative orientations of Dy and Fe moments, occurs with the variation of temperature and/or magnetic-field.

With the primary motivation of confirming the sign-crossover feature in MCE, we have also obtained $C(T)$ curves in zero field and in a high field (figure 5a). The features due to magnetic ordering in the vicinity of respective $T_C$ are masked by large phonon contribution. It is to be noted that, for Dy sample, a peak gradually develops at $T_o$ with increasing $H$. The change in entropy obtained for $H$= 50 kOe is plotted in figure 5b. The observed features (normal and inverse MCE region, and sign-crossover temperature region) are qualitatively similar to that obtained from $M(H)$ data, described above (see figures 4a and 4b). In particular, we would like to emphasize on the following observation, supporting those observed from $M(H)$ data: Reentry into another 'inverse MCE regime' in the range below about 15 K for Dy sample (see inset-I of figure 4b). Non-observation of this 're-entry' phenomenon in the entropy data derived from $C(T)$ for Gd case could be due to sensitivity to measurement conditions as mentioned earlier.

### C. Temperature dependence of magnetoresistance and its relation to magnetocaloric effect

We have carried out MR studies with the intention of tracking MCE behavior, as both these properties depend on the changes in the magnetic configuration due to the application of magnetic field, which has been amply demonstrated in the past literature [see, for instance, Refs 13-14]. Figure 6a show $\rho(T)$ in zero field as well as in 50 kOe for the virgin state of the specimens (that is, obtained while warming the sample after cooling the sample in zero field to desired temperature). As expected, in the zero-field curve, there is a sudden change in slope at respective $T_C$ due to the loss of spin-disorder contribution at this magnetic transition. This kink is smeared out in the 50 kOe curve. There is a peak in MR [defined as $(\rho(H)-\rho(0))/\rho(0)$] at the respective $T_C$ as expected, and the sign of MR (see figure 6b) in a large $T$-range around $T_C$ is negative, which is consistent with net ferromagnetism. A notable observation is that, for Dy sample, there is a sign-crossover at $T_0$. No such sign-crossover could be observed for the Gd sample in the virgin state in the entire $T$-range of investigation (see Fig. 6b), as though compensation point does not exist for the measurement conditions of this technique. All these observations (on sign-crossover) could be verified by performing isothermal MR studies as well. This establishes that thermodynamic and transport properties track moment-configurations. However, sign-reversal at $T^*$, observed in $\Delta S$, is not seen in MR, as though MR data can not distinguish between inverse and normal MCE regimes. We consider that this is an important conclusion.

### D. Nature of hysteretic curves in isothermal magnetization and magnetoresistance

We have also compared hysteresis loops of $M(H)$ and MR$(H)$ at low temperatures (Fig. 7 and Fig. 8). For instance, for Dy sample at 2 K, as mentioned earlier, there is a change in the slope of $M(H)$ curve around 30 kOe in the virgin curve. As the field direction is reversed, there is a gradual decrease deviating from the virgin curve below about 50 kOe (see Fig. 7a); on entering the negative field direction, the -30 kOe-transition becomes discontinuous; on reversal of the field direction to positive zone, the sharpness of the transition near 30 kOe persists. Thus, 'annealing' by field cycling has a profound effect on making 'continuous' transitions in magnetization, presumably arising out of crystallographic defects, into 'discontinuous' ones. With increasing temperature, the hysteresis loop in $M(H)$ gets gradually narrower, the transition field reduces and the transition gets relatively more broadened (see figure 7a for the behavior at 25, 60 and 125 K). Similar observations have been reported by us in the past in some other rare-earth intermetallics [15]. Now, looking at the hysteresis loop in MR($H$) (Fig. 8a), following the sharp change in the field-induced transition around 30 kOe, there is a continuous increase in (positive) MR till 140 kOe. The reversal of the field follows the virgin curve till about 50 kOe as in $M(H)$. However, what is interesting is that, on this reversal of the field-direction, at the point of field-induced transition (near -35 kOe), there is a dramatic drop in the resistivity, with a sign-crossover of MR at this field from 'positive' to 'negative'. On further increase of field (in the negative $H$-direction), there is a gradual reduction in the value of MR and there is another sign-crossover around -80 kOe. On decreasing the field, the curve is hysteretic in the range (about) -35 to 35 kOe. If one looks at the figure 8b, the hysteresis loop is highly asymmetric. This could be due to extreme sensitivity of the scattering process to subtle changes induced by magnetic-field in the relative orientations of the moments of the sublattices. Similar behavior persists at higher temperatures, say at 25K, though the hysteresis loop gets narrower with increasing temperature. Even when the hysteretic behavior almost disappears (say, at 60 K and 125 K, see Fig. 8 ), the above-described sign-crossover & asymmetry persists.

Now, we compare MR($H$) and M($H$) behavior of the Gd compound (see Fig. 7b and 8b). It could be seen clearly that the observed hysteresis loop in $M(H)$ is relatively narrower at all temperatures (see figure 7b for the $M(H)$ behavior at 2, 25, 60 and 125 K) while compared with that of Dy sample, with no tendency for $M$ to saturate even at high magnetic fields. Further, MR($H$) exhibits smaller hysteresis loops at low fields as in $M$ (Fig. 8b). Noteworthy findings are: Unlike in Dy sample, MR is negative in the positive quadrant and the change is rather sharp at low fields tracking magnetization behavior (figure 7b); however, beyond certain field range (about >30 kOe), it exhibits a gradual decrease in magnitude, say, at 2 and 25 K, as though there is a tendency for a sign-crossover at further higher fields. As one enters negative quadrant, the sign of MR tends to be positive, however with reduced values while comparing with those in the positive quadrant for a given field. The values of MR is close to zero at 25 K in the range -50 to -150 kOe. Thus, a clear asymmetry of MR($H$) plots is observed with respect to the zero field even in this case. As in the case of Dy sample, asymmetry with a sharp change around zero-field is present even at higher temperatures (see the curves for 40 K and 125 K in Fig. 8).

It is worth pointing out that, in the case of disordered broadened first-order transitions, one usually observes a 'butterfly' shaped MR(H) curve, in which case the virgin curve lies outside the envelope curve [16]. Therefore, the distinctly different hysteresis loop behavior reported here may not be a consequence of (disorder-broadened) first-order field-induced transition. We think that observed asymmetric MR($H$) curve behavior is interesting, as it is qualitatively similar to that reported for antiferromagnetic tunnel junctions [17]. In these tunnel junctions, the presence of antiferromagnetic region pinned to soft ferromagnetic regions was proposed to be responsible for this asymmetry. It is not clear whether similar reasoning can be given for our cases as well triggered by site disorder.

## IV. CONCLUSIONS

We have compared the behavior of magnetocaloric effect (as measured by isothermal entropy change) and magnetoresistance in the two-magnetic-sublattice systems, GdFe$_5$Al$_7$ and DyFe$_5$Al$_7$. While we observe a peak and a sign-crossover at the respective $T_C$s and zero magnetization points respectively in MCE and MR, there are additional sign-reversals in MCE for Dy compound indicating the existence of multiple inverse-MCE regimes, which is a novel finding. In addition, while MR generally tracks MCE, there are some deviations: (i) There is no sign reversal in MR, as one moves away from normal MCE regime to inverse MCE regime, as the temperature is lowered just after entering magnetically ordered state in both cases. (ii) The MR(*H*) hysteresis loops at low temperatures are highly asymmetric with respect to the origin and there are sign crossovers. It is interesting that such asymmetric MR curves reported for antiferromagnetic tunnel junctions can be seen even in polycrystalline bulk form [18]. Clearly, the scattering process as measured by transport properties appears to be more sensitive than MCE to detect subtle changes in the relative magnetic moment orientations in such multiple magnetic sublattice systems.

**Fig. 1:** X-ray diffraction patterns of RFe$_5$Al$_7$ alloys **(a)** R= Gd, and **(b)** R= Dy. The indexing of diffraction peaks was carried out with the help of Powdercell software. The Rietveld analysis of these patterns show single phase formation and the obtained parameters from the Rietveld refinement are shown in respective plots.

**Fig. 2:** Zero-field-cooled and field-cooled magnetization curves of RFe$_5$Al$_7$ (R= Gd and Dy) taken with 100 Oe and 5 kOe. The lines through the data points serve as guides to the eyes.

**Fig. 3:** Isothermal magnetization (virgin) curves for RFe$_5$Al$_7$ **(a)** R= Gd and **(b)** R= Dy at selected temperatures. The lines through the data points serve as guides to the eyes.

**Fig. 4:** Temperature dependence of isothermal entropy change ($\Delta S$), obtained from isothermal magnetization data using Maxwell relation, are plotted for RFe$_5$Al$_7$, **(a)** R= Gd and **(b)** R= Dy alloys. In the insets, the data at low temperatures are plotted as a function of temperature and also as a function of magnetic field. The lines through the data points serve as guides to the eyes.

**Fig. 5:** Temperature dependence of (a) heat-capacity divided by temperature and (b) isothermal entropy change calculated from heat capacity data for RFe$_5$Al$_7$ (R= Gd and Dy). Inset in (b): The entropy change in the lower temperature region is shown in an expanded form. The lines through the data points serve as guides to the eyes.

**Fig. 6: (a)** Temperature dependence of electrical resistivity ($\rho$) for RFe$_5$Al$_7$ (R= Gd and Dy) alloys. The magnetoresistance (MR) values derived from these plots are plotted in **(b)**. The MR data show higher values nearer to the magnetic ordering temperature for both alloys. The lines through the data points serve as guides to the eyes.

**Fig. 7:** Isothermal magnetic hysteresis curves for RFe$_5$Al$_7$ (R= Gd and Dy) alloys at selected temperatures (2, 25, 60 and 125 K a). Insets show the plots in an expanded region in the low-field range to highlight hysteretic behavior. The lines through the data points serve as guides to the eyes. The arrows and numericals on the curves are placed to show the direction of applied magnetic field variation.

**Fig. 8:** Isothermal MR curves for RFe$_5$Al$_7$ (R= Gd and Dy) at selected temperatures (2, 25, 40, and 125 K). The lines through the data points serve as guides to the eyes. The arrows and numericals are placed on the curves to show the direction of applied magnetic field variation.

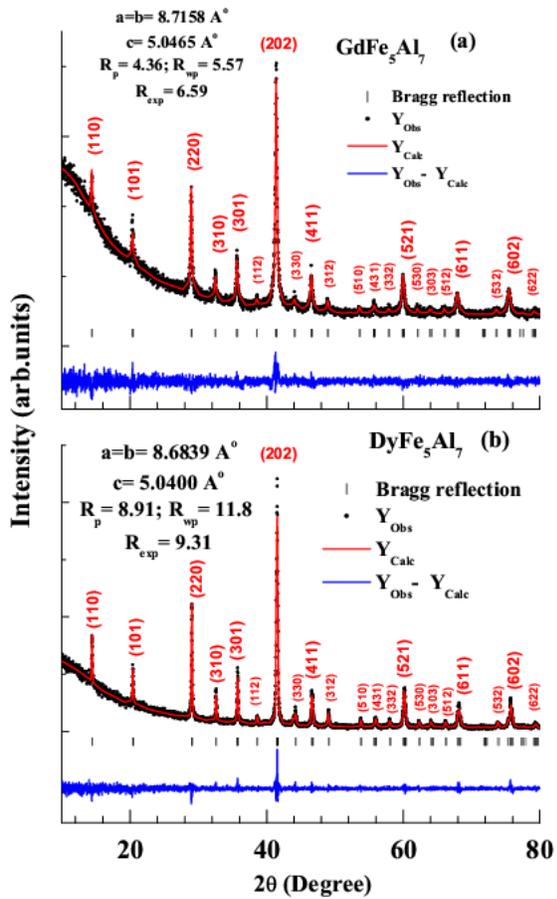

Figure 1

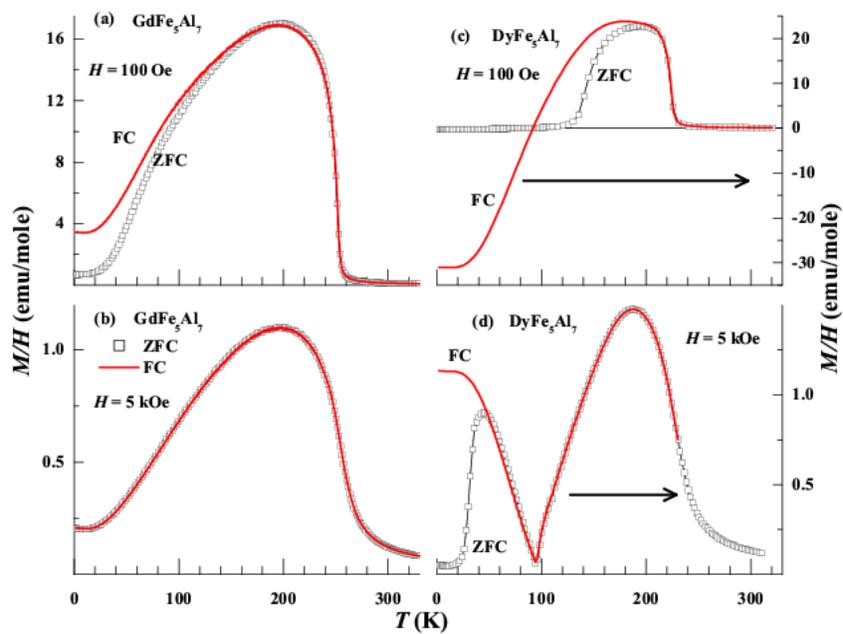

Figure 2

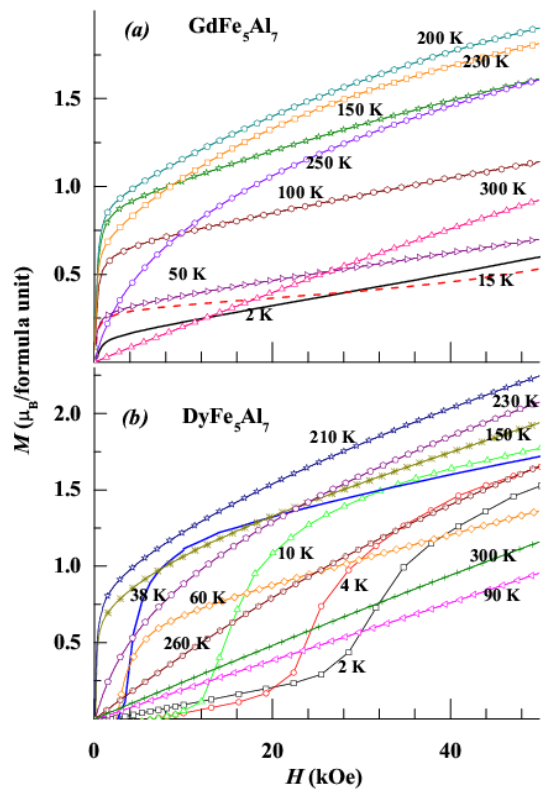

Figure 3

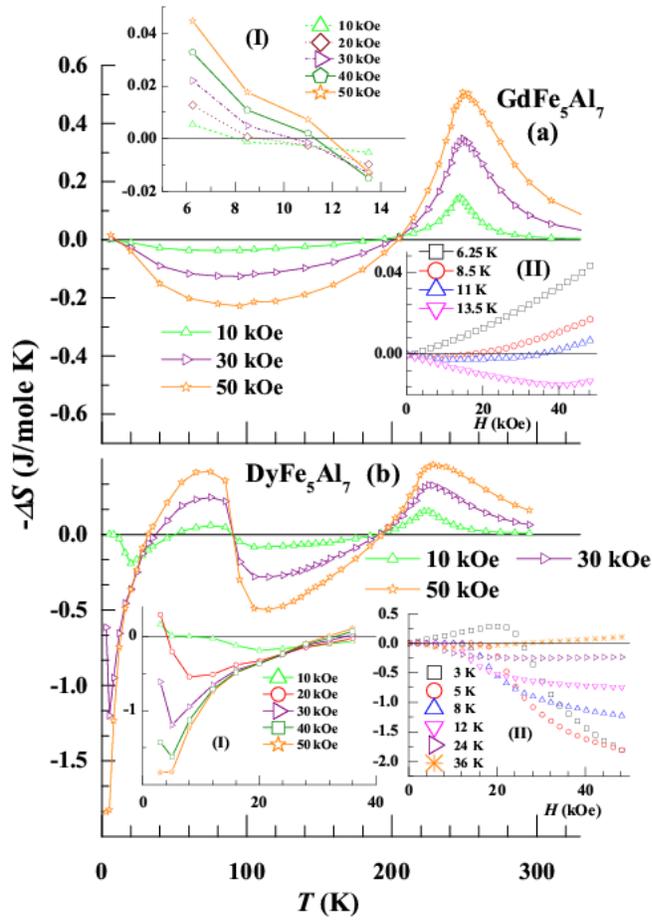

Figure 4

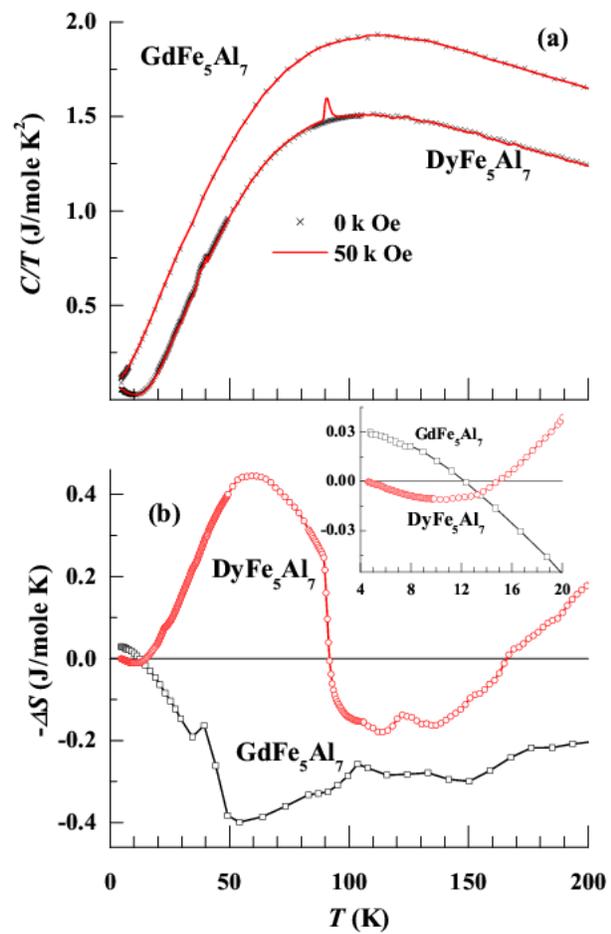

Figure 5

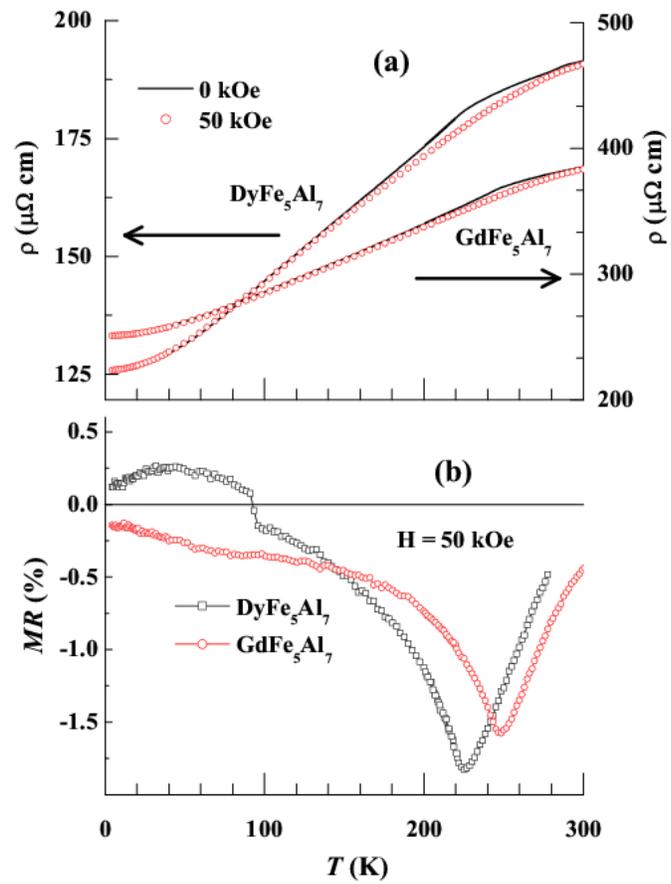

Figure 6

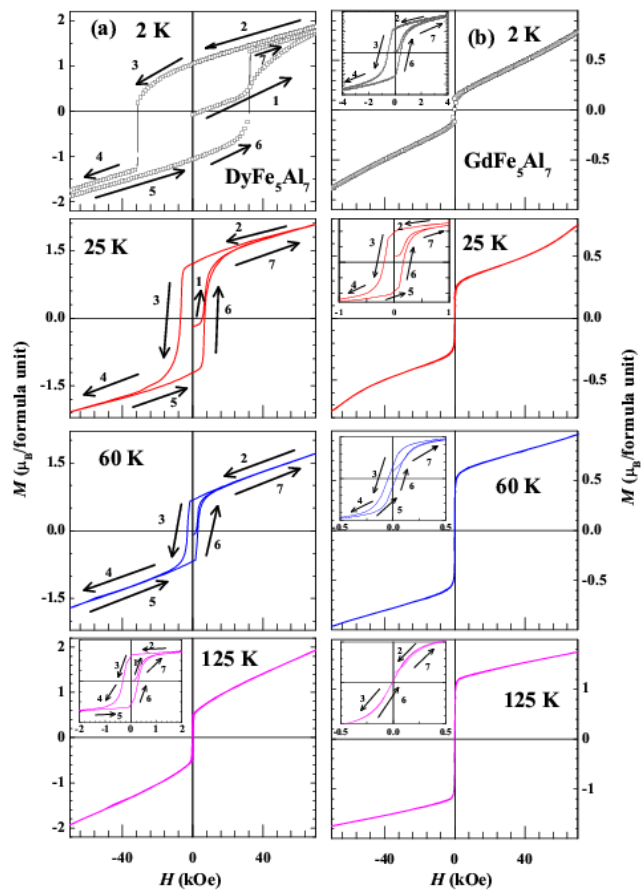

Figure 7

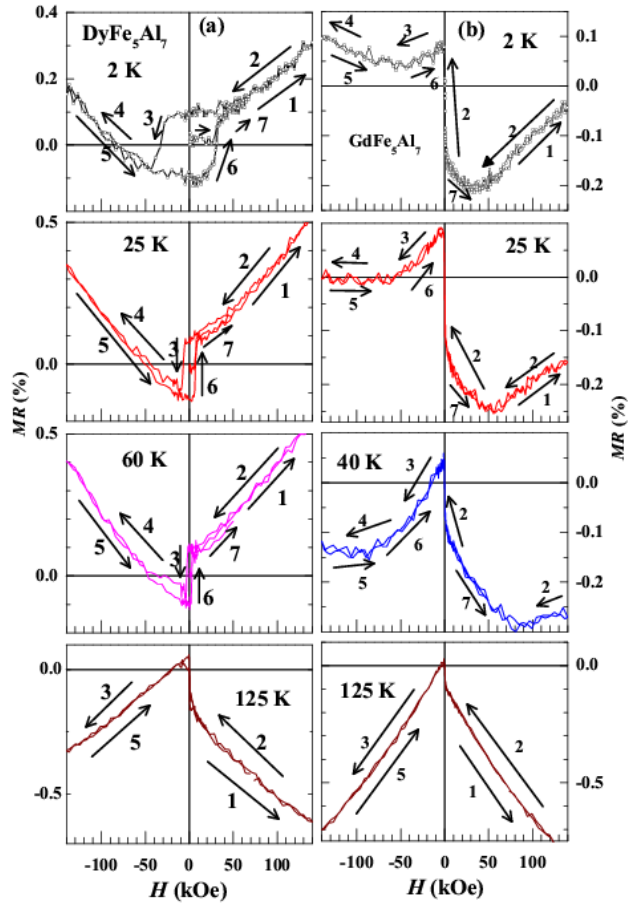

Figure 8